\def\references{\subsection*{\normalsize\bf References}
\bgroup\parindent=0pt\parskip=\itemsep
\def\refpar{\par\hangindent=1.2em\hangafter=1}}
\def\endreferences{\refpar\egroup}

\def\@biblabel#1{\relax}
\def\@cite#1#2{#1\if@tempswa , #2\fi}
\def\reference{\relax\refpar}
\def\@citex[#1]#2{\if@filesw\immediate\write\@auxout{\string\citation{#2}}\fi
\def\@citea{}\@cite{\@for\@citeb:=#2\do
{\@citea\def\@citea{,\penalty\@m\ }\@ifundefined
{b@\@citeb}{\@warning
{Citation `\@citeb' on page \thepage \space undefined}}%
{\csname b@\@citeb\endcsname}}}{#1}}

\documentstyle[11pt]{article}
\begin{document}

\def\slantfrac#1#2{\hbox{$\,^#1\!/_#2$}}
\def\onethird{\slantfrac{1}{3}}
\def\gamt{$\bra \gamma_{\scriptscriptstyle T}\ket$}
\def\h1{$h^{-1}$\/}
\def\bra{\mathopen{<}}
\def\ket{\mathclose{>}}
\def\lefevre{Le\thinspace F\`evre~}
\def\miralda{Miralda-Escud\`e~}
\def\ms0440{MS\thinspace 0440$+$0204}
\def\ms1054{MS\thinspace 1054$-$03}
\def\etal{et~al.~}
\def\kmpersec{km~s$^{-1}$~}
\def\ergpersec{erg~s$^{-1}$~}
\def\ergspersec{erg~s$^{-1}$~}
\def\ergspersqcmsec{erg~s$^{-1}$~}
\def\approx{$\,\sim\,$}
\def\dash{$\,-\,$}
\def\gequal{$\,\geq\,$}
\def\lequal{$\,\leq\,$}
\def\gthan{$\,>\,$}
\def\lthan{$\,<\,$}
\def\equal{$\,=\,$}
\def\equals{$\,=\,$}
\def\msun{M$_\odot\,$}
\def\lsun{L$_\odot\,$}
\def\la{\mathrel{\hbox{\rlap{\hbox{\lower4pt\hbox{$\sim$}}}\raise1pt\hbox{$<$}}}}
\def\ga{\mathrel{\hbox{\rlap{\hbox{\lower4pt\hbox{$\sim$}}}\raise1pt\hbox{$>$}}}}

\begin{center}
{\Large\bf
Detection of weak lensing by a cluster of galaxies at {\boldmath $z$\equals}0.83
}
\vskip 10pt
{\large
{\sc G. A. Luppino}\\
\vskip 2pt
{\small Institute for Astronomy, University of Hawaii \\
 2680 Woodlawn Drive, Honolulu, Hawaii 96822, email: ger@hokupa.ifa.hawaii.edu}\\
\vskip 5pt
{\sc and}\\
\vskip 5pt
{\sc Nick Kaiser}\footnote{
Temporarily on leave at Anglo-Australian Observatory, Epping
Laboratory, NSW 2121, Australia}\\
\vskip 2pt
{\small Canadian Institute for Advanced Research and
Canadian Institute for Theoretical Astrophysics \\
University of Toronto, 60 St. George St., Toronto, Canada M5S 1A7,
email: kaiser@cita.utoronto.ca}
}
\end{center}

\begin{center}
Submitted to {\it ApJ}
\end{center}

\vskip 10pt
\begin{center}
{\large\bf Abstract \vspace{-.5em}\vspace{0pt}}
\end{center}
{\quotation
We report the detection of weak gravitational lensing of
faint, distant background galaxies by the
rich, X-ray luminous cluster of galaxies \ms1054
at $z$\equals 0.83.  This is the first measurement of weak
lensing by a bona fide cluster at such a high redshift.
We detect tangential shear at the 5\% - 10\% level over
a range of radii $50'' \la r \la 250''$ centered on the optical
position of the cluster.
Two-dimensional mass reconstruction using galaxies with
$21.5 < I < 25.5$ shows a strong peak which coincides with the
peak of the smoothed cluster light distribution.
Splitting this sample by magnitude (at $I = 23.5$) and
color (at $R-I = 0.7$), we find that the brighter and redder subsamples
are only very weakly distorted, indicating that the faint blue
galaxies (FBG's), which dominate the shear signal, are relatively more
distant. The derived cluster mass is quite sensitive to the $N(z)$ for the
FBG's.  At one extreme, if all the FBG's are at $z_s = 3$, then
the mass within a $0.5h^{-1}$Mpc aperture is
$(5.9 \pm 1.24)\times 10^{14}$\h1 $M_\odot$, and
the mass-to-light ratio is $M/L_V = 350 \pm 70 h$ in solar units.
For $z_s = 1.5$ the derived mass is $\sim$70\% higher and
$M/L \simeq 580 h$.  If $N(z)$ follows the no evolution
model (in shape) then $M/L \simeq 800h$, and if all the FBG's lie at
$z_s\la 1$ the required $M/L$ exceeds $1600h$. These data provide clear
evidence that large, dense mass concentrations existed at early epochs;
that they can be weighed efficiently by weak lensing observations;
and that most of the FBG's are at high redshift.

\vskip 5pt
\begin{tabbing}
Subject headings:~~\= \kill 
{\it Subject headings:} \>cosmology: observations  ---  gravitational
lensing
--- dark matter --- \\
\> galaxies: photometry --- galaxies: distances and redshifts  ---\\
\> galaxies: clusters: individual (MS\thinspace 1054$-$03).\\

\end{tabbing}
\par
\vskip 1pt
}

\par\noindent
{\large\bf 1. Introduction}

The technique of weak gravitational lensing has emerged as a powerful probe both
of clusters of galaxies and of the faint blue galaxy [FBG] population.
Most weak lensing observations to date have
concentrated on  low and intermediate redshift clusters
($z$\approx 0.2--0.4); for example  
A1689 at $z$\equals 0.18 (Tyson, Valdes \& Wenk 1990; 
Tyson \& Fischer 1995; Kaiser, Broadhurst, Szalay and Moller, 1996), 
A2218 at $z$\equals 0.18 (Squires \etal 1995), 
MS1224$+$24 at $z$\equals 0.33 (Fahlman \etal 1994),
A370 at $z$\equals 0.375 (Kneib \etal 1994), and Cl\thinspace 0024$+$17 at
$z$\equals 0.39 (Bonnet \etal 1994).
Clusters in this redshift range are sufficiently 
far away that they can be imaged
efficiently with existing 2048$^2$ pixel CCD detectors, and yet are 
close enough that
the derived mass is little affected by uncertainty in the redshifts
of the faint lensed galaxies.

Observing lensing by high-redshift ($z$\gthan 0.7) clusters is
more difficult, since for a lens of a given mass the
distortion tends to weaken with increasing lens redshift,
especially as the lens redshift approaches that of the sources.
However, this dependence of the distortion strength on the
observer-lens-source geometry potentially provides a powerful constraint
on the redshift distribution $N(z)$ of faint galaxies. 
If the majority of these lie at high redshift ($z$\gthan 2, say),
then we should
see strong distortion for even the most distant ($z$\approx 1) clusters, but
if the majority of faint galaxies lie at or below $z$\approx 1, then
the distortion should fall rapidly as the cluster redshift approaches unity.
In this way, one can constrain $N(z)$ at much fainter magnitudes ($I$\gthan 24)
than are accessible
by spectroscopic surveys, even with the new generation of 8--10\thinspace m
telescopes.   

Smail \etal (1994) tried this experiment
by looking for weak lensing in
three clusters covering a wide range of
redshifts ($z$\equals 0.26, $z$\equals 0.55 and $z$\equals 0.89).
A clear lensing signature was seen in the $z$\equals 0.26 cluster, and
a somewhat weaker signal in the $z$\equals 0.55 cluster, 
but none was seen in the highest redshift
cluster, Cl\thinspace 1603$+$43 at $z$\equals 0.89, 
suggesting that the majority of
FBGs with $I$\lthan 25 were at $z\la 1$.
However, an alternative interpretation is that 
Cl\thinspace 1603$+$43 is simply not massive enough to produce a 
measurable shear signal. This is not implausible since 
this cluster was optically selected (Gunn \etal 1986), and has an
X-ray luminosity of only $L_x$\approx 1$\times$10$^{44}$ erg s$^{-1}$;
(Castander \etal 1994), as compared to the two lower-redshift
clusters which both have $L_x$\gthan 10$^{45}$ erg s$^{-1}$.
Of course, Smail \etal had little to choose from.
When they performed their observations, there were no known clusters at
$z$\gthan 0.7 with X-ray luminosities comparable to the richest and
brightest low-redshift clusters, and the small number of
high-$z$ clusters then known were mainly optically detected 
(e.g. Gunn \etal 1986; Couch \etal 1991).
Recently, however, several new, high-redshift clusters have been discovered 
as the optical counterparts to previously-unidentified
{\it Einstein} Extended Medium Sensitivity Survey
(EMSS) X-ray sources
(Gioia \etal 1990; Gioia \& Luppino 1994).
The most distant of these, \ms1054 at $z$\equals 0.83, is extremely rich 
and has an X-ray luminosity
an order of magnitude higher than Cl\thinspace 1603$+$43 (Luppino \& Gioia 1995),
suggesting it may be a potent gravitational lens.

In this paper, 
we report the detection of weak gravitational lensing by \ms1054.
Our observations and data reduction are outlined in \S 2,
the cluster properties are described in \S 3.
In \S 4, we apply weak lensing analysis, and in 
\S 5, we discuss the implications 
of our observations for cosmological structure formation models,
and for the constraining the redshift distribution of the faint
background galaxies.

\par\noindent
{\large\bf 2. Observations and data reduction}

Optical $R$ and $I$-band images of \ms1054 were obtained with the UH 2.2m 
telescope on the nights of 19 Feb 1993 and 11-13 Jan 1994. A thinned Tek
2048$^2$ CCD was mounted at the f/10 
RC focus resulting in a scale of 0$''$.22/pixel and a field 
of view of 7$'$.5$\times$7$'$.5 (physical scale
$1.86 h^{-1}$Mpc at $z = 0.83$). The total 
exposure times were 7200\thinspace s 
and 21600\thinspace s in $R$ and $I$ respectively.
The individual images in each filter were first de-biased and then
flattened using a median of all the CCD frames taken in that filter 
(including the cluster images which made up \approx $\onethird$  of the 
total number of frames).  Low spatial frequency
residual sky fluctuations were then removed by subtracting a highly
smoothed image determined from the troughs of the minima in the images.  
Registration was performed using $\sim50$ moderately
bright stars, and the images were then transformed to a common
coordinate system (with bi-linear interpolation).  The stack of transformed
images was then summed with cosmic-ray rejection and using appropriate
weights (the cosmic-ray rejection being done in such a way as to ensure
that the effective psf for the stars was the same as for the faint
objects).  The seeing in the resulting $R$ and $I$ images was 1$''$.14 
and 0$''$.97 FWHM respectively. Photometric calibration was performed using 
the standard stars of Landolt (1993).
The variation in extinction between the $I$-band
images was very small, as was
also the case for all but three of the $R$-band images.
The 1$\sigma$ surface brightness limits of the 
summed $R$ and $I$ images are 27.9
mag arcsec$^{-2}$ and 27.8 mag arcsec$^{-2}$ respectively.

In order to detect  the faint objects we used the
algorithm of Kaiser, Squires \& 
Broadhurst (1995 [KSB]). 
This provides a catalog with accurate positions but crude
size and magnitude information.  We then used this catalog to mask the
summed images and thus determine and subtract the small residual positive
bias in the images left by the local sky subtraction, and
we then applied photometric analysis to obtain refined sizes, magnitudes etc.
The resulting
catalog contained some noise peaks as well as detections of
groups of objects.  These were removed
by limiting the catalog at 5-sigma detections and removing abnormally
small and large objects. We also rejected a small number of objects
with high eccentricity to obtain final
catalogs containing $N_I = 2718$ and $N_R
= 1822$ objects, corresponding to about $1.7 \times 10^5$ and $1.2 \times 10^5$
objects per square degree.  Nearly all the objects detected in the $R$-band were
also detected in $I$. 
The $I$-magnitudes
were determined using a large aperture $r_{\rm ap} = 3 r_g$,
where $r_g$ is the smoothing scale at which the object was
detected, and typically overestimate total magnitudes by $\la 0.1$ mag.  

\par\noindent
{\large\bf 3. Cluster properties}

\begin{figure*}[b]
\vspace{.2in}
\begin{minipage}[t]{4.8in}
\vspace{4.3truein}
\includegraphics{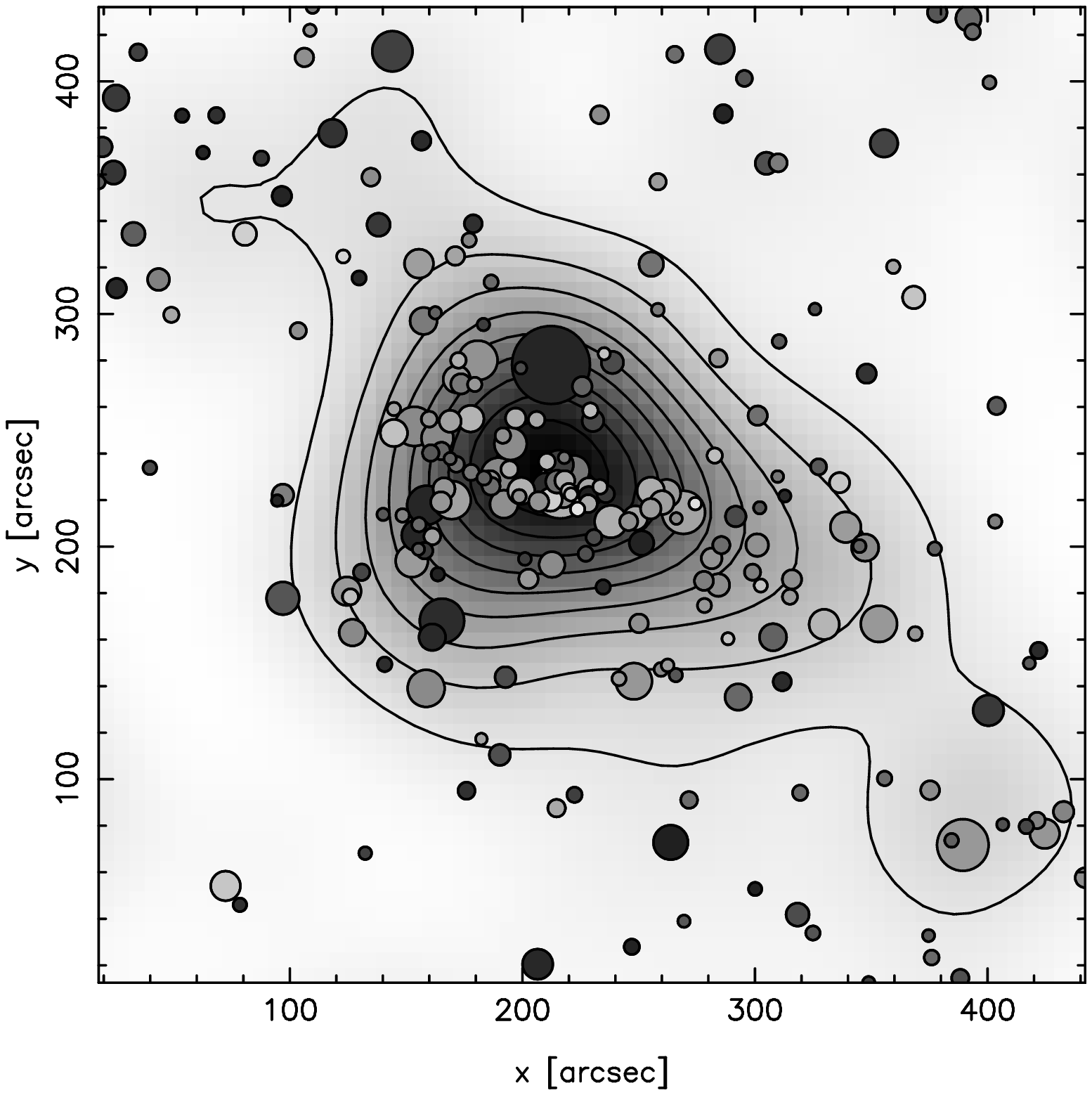}
\end{minipage}
\hfill
\begin{minipage}[t]{2.6in}
{\small{\sc Figure 2.}
Spatial distribution of red galaxies (including the
sequence of cluster galaxies with $R-I \simeq 1.5$).  The size of each
circle is proportional to the brightness of the galaxy (in $I$), and the
shading indicates
the color on a scale of $R-I = 1.9$ (white) to $R-I = 1.1$ (black).
The underlying gray-scale is the $I$-band surface brightness
smoothed with a $35''$ gaussian filter.
\label{fig2} 
}
\end{minipage}
\end{figure*}

\ms1054 is an extraordinary object.  
It is by far the richest and most X-ray luminous high-redshift ($z$\gthan 
0.7) cluster
known, and is among the richest clusters known at any redshift. A true
color image centered on the $I$\equals 19.3
brightest cluster galaxy (BCG) is shown in figure 1 [Plate 1];
the cluster is easily identified as
the horizontal swath of red galaxies in the center of the frame.
Figure 2
shows the location, I-magnitude and color of all the 
non-stellar objects with $I < 24.5$ and with colors in the range
$1.93 > R-I > 1.1$ which brackets the color of the cluster galaxies.  
The total magnitude for all of 
the galaxies contained within a $1'$ aperture (physical
scale of $\simeq 0.25 h^{-1}$Mpc for $q_0 = 0.5$)
centered on the brightest cluster galaxy is $I = 16.5$.
Converting the observed $I$-band magnitude to a rest-frame solar
luminosity $L_{V\odot}$ using the relation 
$M_V$\equals$I-5\,\log{[(1+z_l)^2\,D_l]}-25+(V-I)_{\rm o}-K(z)$ with
the K-correction $K(z)$=0.85, $(V-I)_{\rm o}$\equals 1.3, and
$M_{V\odot}$\equals +4.83 we obtain 
$L(<\,0.25h^{-1}\,{\rm Mpc}) = 1.19 \times 10^{12} L_{V\odot}$, which includes
a $\sim$ 15\% contribution
from the bright foreground galaxy lying $\sim 1'$ to the north of the cluster center.
For a $0.5 h^{-1}$Mpc
aperture we find $L(<0.5) = 2.0 \times 10^{12} L_{V\odot}$. 
The number of galaxies with $I < 22$ counted within the same apertures are
$N(<0.25) = 49$ and $N(<0.5) = 82$, which represent an excess
over the background of about 44 and 67 galaxies respectively, making this
at least a richness class 4 cluster (Bahcall 1981).

Although \ms1054 is clearly very X-ray luminous
($L_{\scriptsize 0.3-3.5\,{\rm keV}}$\equals9.3$ \times 10^{44}$
$h_{50}^{-2}$ erg s$^{-1}$), the actual X-ray flux is quite low
because the cluster is so distant, and consequently little can be said
about its X-ray properties at the present time. 
\ms1054 was unresolved in the {\it
Einstein} IPC with only 107.9$\pm$12.8 counts in an 18\thinspace ksec
exposure, corresponding to a flux of 
$f_x$\equals 2.11$\times$10$^{-13}$ erg cm$^{-2}$ s$^{-1}$ 
(Henry \etal 1992). The flux was
converted to a luminosity assuming a 6 keV thermal spectrum and
correcting for extended emission as outlined in Gioia \& Luppino
(1994). An {\it ASCA} spectrum has recently been obtained,
 and a preliminary analysis
indicates the cluster has a high X-ray temperature (Donahue, private communication).
ROSAT HRI observations are scheduled.

\par\noindent
{\large\bf 4. Weak lensing analysis}

The weak lensing analysis involves several steps. 
Object polarizations
$e_{\alpha}$\equals$\{I_{11}-I_{22},2\,I_{12}\}/(I_{11}+I_{22})$ 
were formed from the the quadrupole moments 
$I_{ij}=\int d^2\theta\, W(\theta) \theta_i \theta_j f({\boldmath
\theta})$ where $f$ is the flux density and 
$W(\theta)$ is a gaussian weighting function
matched to the size of the galaxy.
We then extract a sample of moderately bright stars which have non-zero
polarization due to anisotropy of the point spread function, fit
a low order polynomial model for the psf variation across the field,
and then correct the galaxy polarizations for all the objects to what they would
have been for perfectly circular seeing as described
in KSB.  These $e_\alpha$ values should now be equal to the random
intrinsic values plus a small coherent shift which is proportional
to the gravitational shear 
$\gamma_\alpha = {1\over 2}\{\phi_{,11}-\phi_{,22},2\phi_{,12}\}$
where $\phi$ is related to the dimensionless surface mass density
by $\kappa=\Sigma/\Sigma_{crit}={1\over 2}\nabla^2\phi$ and where the
critical density
$\Sigma_{crit}^{-1}= {4\pi G c^{-2}} {D_l D_{ls} D_s^{-1}}
= {4\pi G c^{-2} D_l\beta},$ with
$\beta \equiv D_{ls}/D_s$ ($ = [1-D_l(1+z_l)/D_s(1+z_s)]$
for $\Omega=1$).

The next step is to calibrate the relation between the polarization and
the shear. Previously, this has been done by 
artificially shearing deep HST images
to simulate lensing and convolving with a gaussian seeing disk (KSB).
Here we have used a slightly different approach. We artificially
shear the actual I-band image (which is equivalent to shearing
the image as it would appear from space, but then convolving with a 
slightly anisotropic psf),
and then correct the galaxy polarizations using the sheared stars.
This is equivalent (for small shear at least)
to shearing the image before seeing and then smoothing
with a circular psf, and the shear polarisability is then
just equal to the change in the polarization divided by the
applied shear, $P_\gamma = d e / d \gamma$.  
The individual $P_\gamma$ values are rather noisy for
the faintest objects, but the mean polarisability varies smoothly in the
way expected with radius, and should be adequate to determine the
appropriate calibration factor $\langle P_\gamma \rangle$
for each of the subsamples we will
construct.  This new approach gives results which agree very
well with those from the previous method using HST images (KSB), but is
more convenient here.  We now have a fair estimate of the shear 
$\hat \gamma_\alpha = e_\alpha / \langle P_\gamma \rangle$
for each galaxy --- albeit a rather noisy one ---
which we now analyze in a number of different ways, and also using 
various subsamples.

\begin{figure*}[t]
\vspace{4.2in}
\includegraphics{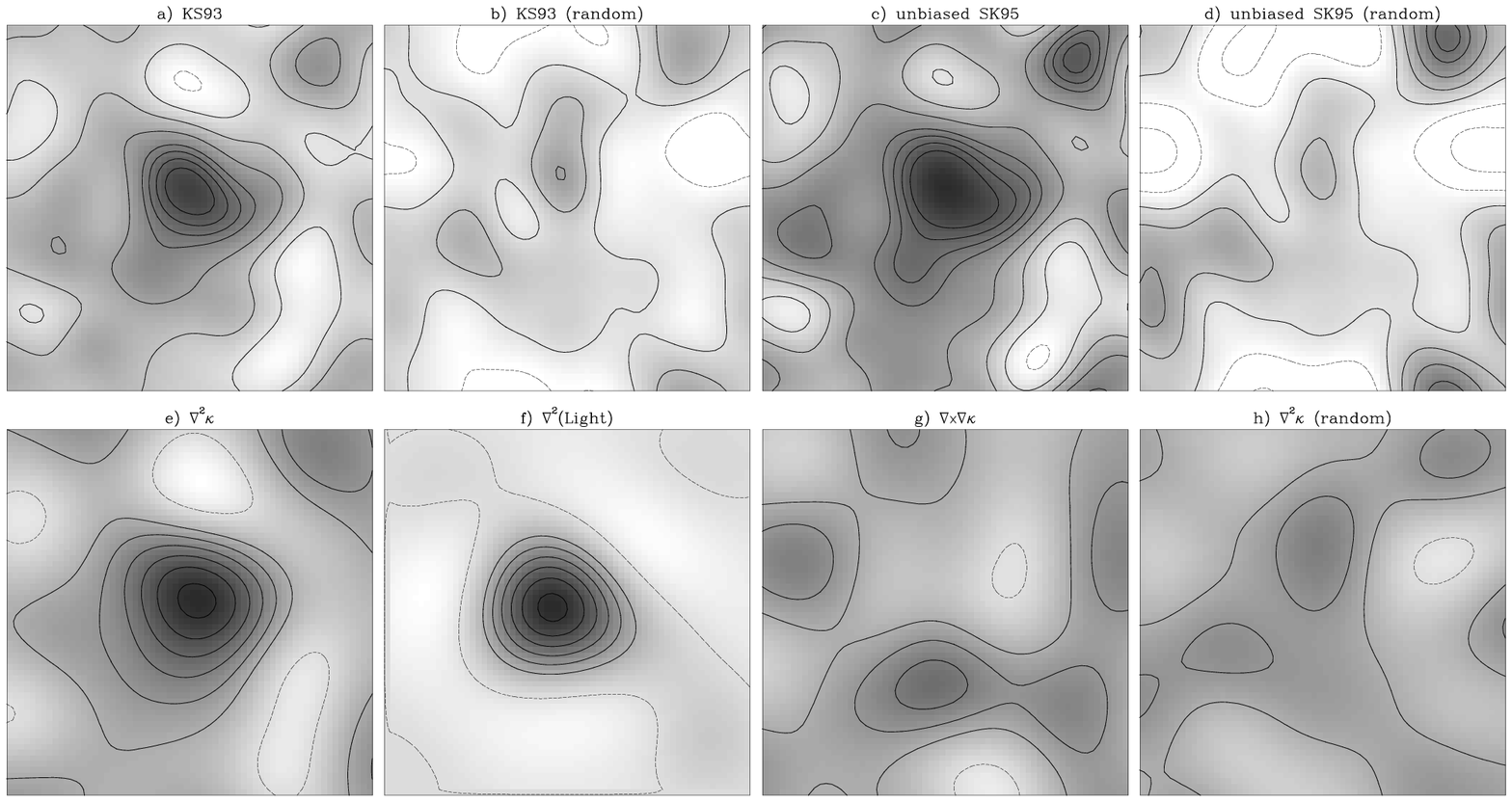}
\begin{center}
\begin{minipage}{6.5in}
\par\noindent
{\small{\sc Figure 4.}
The top four panels show the result of two different mass
reconstruction algorithms:
a) the original KS93 method and c) the new, unbiased
`regularized maximum likelihood' technique of Squires \& Kaiser
(1996).
While the KS93 method is susceptible to a slight
negative bias at the edge of the field (Schneider 1995),
it appears that in this case any bias that might be present is small.
Panels b) and d) are reconstructions using a catalog
in which the galaxies were assigned normally distributed random
shear values
with rms (per component) $\gamma_\alpha = 0.6$, and which indicate the
expected level of noise in these reconstructions.
The lower four panels contain e)  a smoothed image of $\nabla^2\kappa$
(or equivalently $\kappa$ smoothed with a compensated `mexican-hat'
filter), f) the Laplacian of
the surface brightness (scaled to have the same peak value),
g) an estimate of $\nabla \times \nabla \kappa$ which should
be zero if the shear field is really due to gravity, and h)
a realization of the noise produced by our random catalog.
\label{fig4}
}
\end{minipage}
\end{center}
\end{figure*}

First we define a sample of all faint objects in the $I$ catalog 
having $I > 21.5$ (2395 objects).
No attempt was made to remove stars or cluster galaxies.
This faint galaxy sample can be seen in figure 3 [Plate 2]
as ellipses overlaid on the $I$-band CCD image of the cluster.
Figure 4 shows the result of applying two different inversion
algorithms to recover the dimensionless surface density $\kappa(\vec r)$:
the original Kaiser \& Squires (1993) algorithm [KS93] and the new,
unbiased Squires \& Kaiser (1996) algorithm [SK96].
Massmaps generated by either algorithm (see figs 4a and 4c)
show  strong mass concentrations very close to the
peak of the smoothed lightmap.  Also shown are
reconstructions using the same spatial distribution, but with random
gaussian shear values with $\langle \gamma_\alpha^2 \rangle^{1/2}
= 0.6$ (a value determined from the data as described below).
These mass reconstructions have been smoothed to the same $35''$ gaussian
filter scale as the light. 
Figure 5 [Plate 3]
shows a contour plot (white contours) of the cluster light
superimposed on the mass contours (black contour lines) overlaid on 
the $I$-band CCD image of the cluster field. 

While the relation between the shear (essentially the tidal field) and
$\kappa$ is a non-local one, there is an explicit local expression for
the gradient of the surface density
in terms of the gradients of the shear  (Kaiser 1995), and one can therefore
determine $\nabla^2 \kappa$, the Laplacian of the surface density,
from local shear estimates.  A smoothed
image (filter scale = $70''$) of $\nabla^2 \kappa$ is shown
in figure 4e.  The smoothed Laplacian is just the
surface density convolved with a particular form of `mexican-hat'
smoothing filter --- it is because this filter is `compensated' that
the resulting field does not suffer from the slight bias (Schneider
1995) inherent in the KS93 method, 
and so can be compared directly with the Laplacian
of the surface brightness (figure 4f); clearly these agree in shape and
location very well indeed.

An interesting feature of this kind of analysis is that it provides a powerful
check on whether the distortion we are detecting is really 
due to gravitational lensing.
If instead of the Laplacian $\nabla \cdot \nabla \kappa$
we calculate the curl of the gradient
$\nabla \times \nabla \kappa$, we should then get zero plus
fluctuations due to the random noise in the shear estimates.
What we are doing here is exploiting the fact that while a general
distortion field has two real degrees of freedom, one generated
by gravity has only one, and
we are projecting out two
components of the shear field: one which is excited by 
gravitational lensing and another which is not.
To generate $\nabla \times \nabla \kappa$ rather than
$\nabla \cdot \nabla \kappa$
we simply swap the two components of the shear and change the
sign of one of them (this is equivalent to rotating each object by 45 degrees).
Due to the high symmetry of these operations, one would expect
most (but not necessarily all) artificial sources of distortion to excite
both modes, and so the smallness of the estimate of $\nabla \times
\nabla \kappa$ (visible in figure 4g) provides a non-trivial check of 
the reality of the shear field we detect.
Finally, the amplitude of the noise fluctuations expected are indicated in the
lower right panel of figure 4, and we see no
excess of noise due to artificial sources of
image polarization (such as errors in the registration).

\begin{figure*}[t]
\vspace{5.7in}
\includegraphics{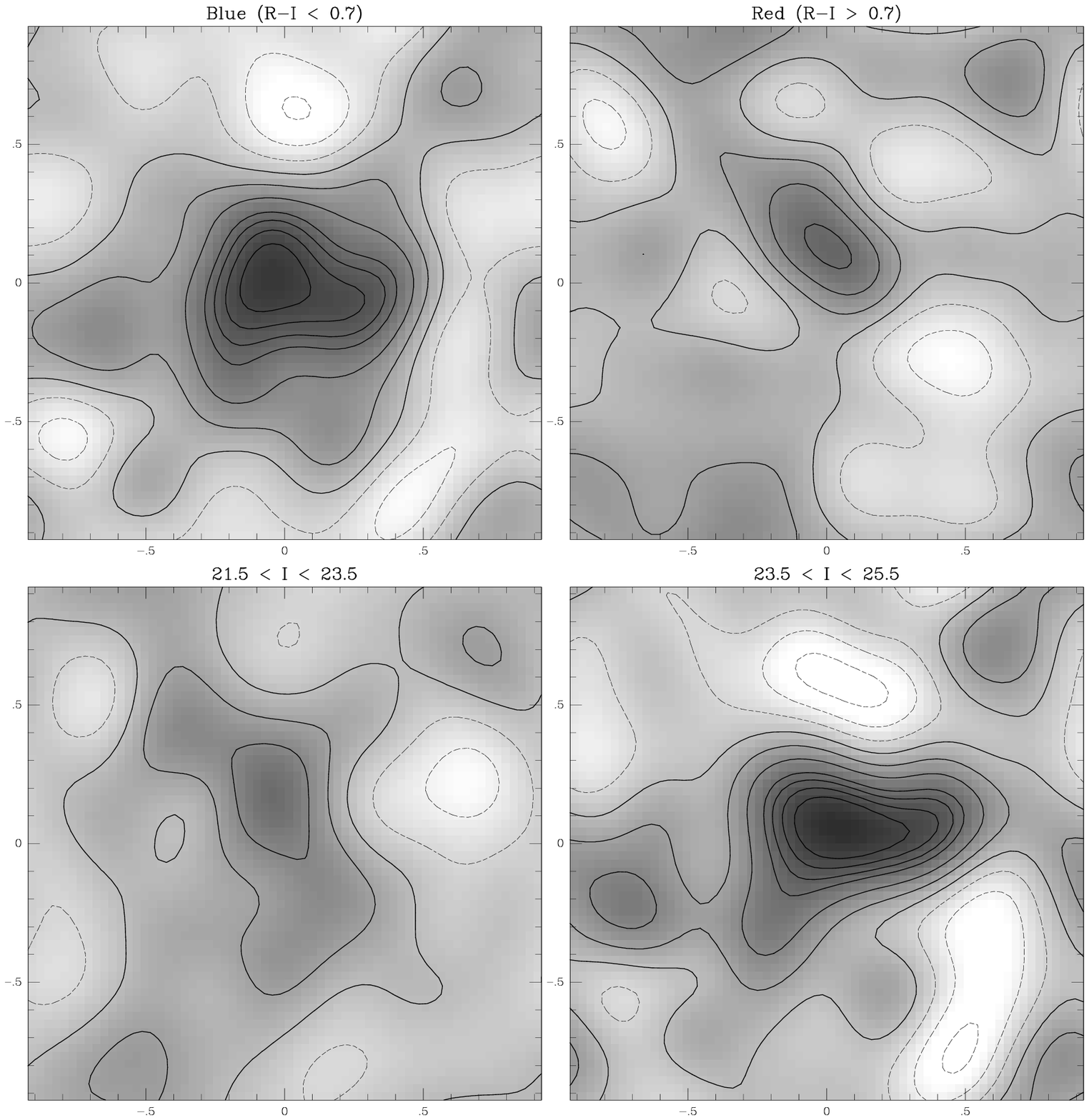}
\begin{center}
\begin{minipage}{6.0in}
\par\noindent
{\small{\sc Figure 6.}
Mass reconstruction from the various galaxy subsamples:
upper left, blue; upper right, red;
lower left, faint-bright galaxies (21.5\lthan$I$\lthan 23.5);
lower right, faint-faint
galaxies (23.5\lthan$I$\lthan 25.5).
The axes are labeled in units of \h1 Mpc.
All four massmaps are displayed with the
same intensity stretch and contour levels.
\label{fig6}
}
\end{minipage}
\end{center}
\end{figure*}

To search for variation in the distance to the background galaxies
we have split the full $I > 21.5$ sample into subsamples by
magnitude (at $I = 23.5$) and color (at $R-I = 0.7$).  The mass
reconstructions for these four (bright, faint, red, blue) subsamples
are shown in figure 6.  The faint and blue
reconstructions are very similar.  They clearly show the cluster, which now appears 
elongated in the same sense as the cluster galaxies, and give a somewhat
higher peak than for the full sample 
(though at a similar 5-sigma level of significance).
The red and bright subsamples, however, show very little sign of the
cluster at all --- as would be expected if the typical redshift of 
these objects is less than or of order unity.  Note that the difference in amplitude
is not a result of different sizes for the background galaxies, as
this is corrected for when we calculate $P_\gamma$; the difference must
reflect a greater distance to the faint and blue objects.

\begin{figure*}[p]
\vspace{8.5in}
\includegraphics{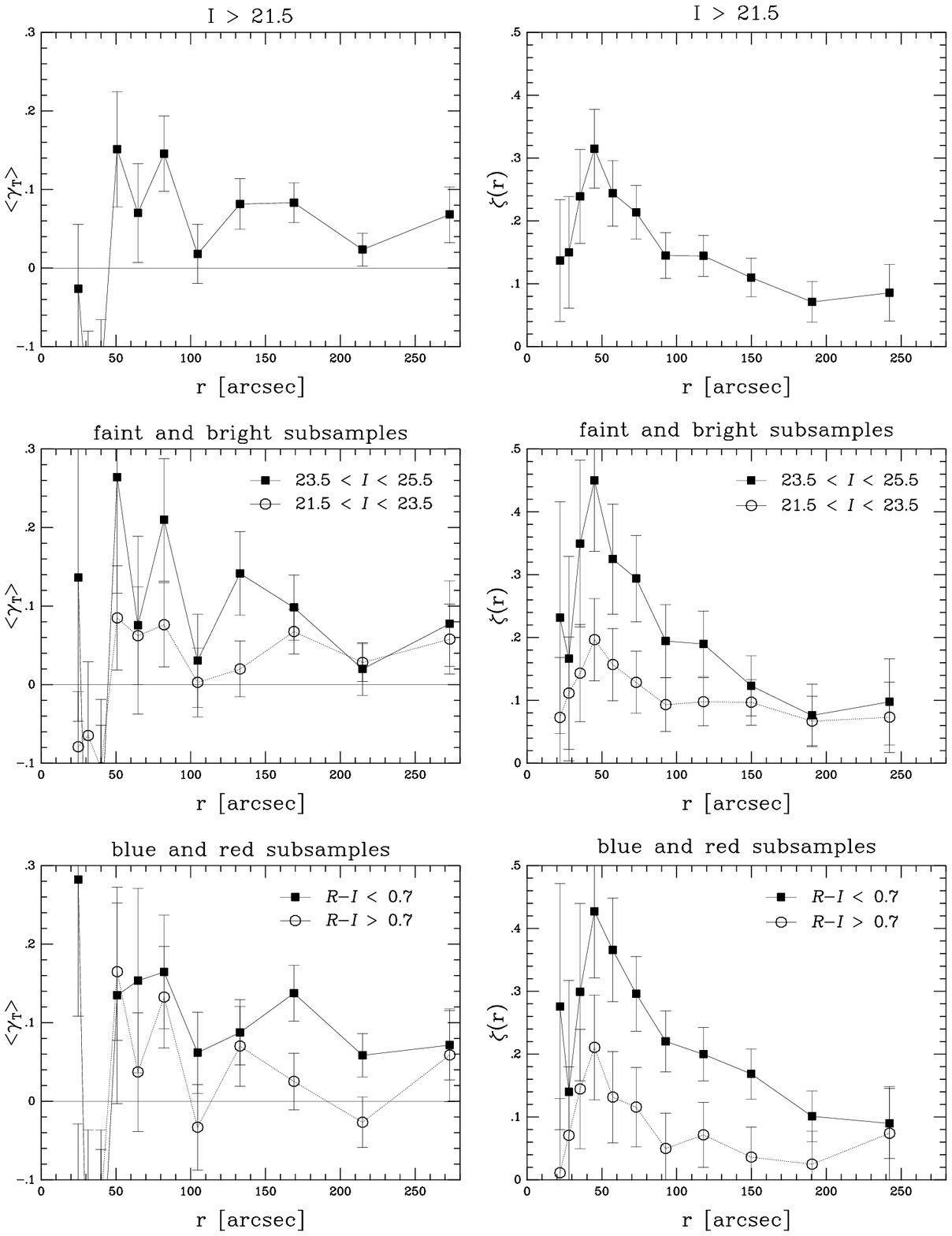}
\par\noindent
{\small{\sc Figure 7.} 
Panels on the left show the tangential shear $\gamma_T$ for the
$I > 21.5$ sample (top); the faint and bright subsamples are shown
as square and circular symbols in the middle panel and the
blue (square) and red (circle) samples are shown in the bottom panel.
The righthand panels show $\zeta(r)$ which provides a lower bound
on $\overline \kappa(r)$.
\label{fig7}
}
\end{figure*}

In addition to the 2D mass reconstruction we have performed
``aperture mass densitometry''.
The statistic
\begin{equation}
\zeta(r) = (1-r^2/r_{\rm max}^2)^{-1}
\int_{r}^{r_{\rm max}} \bra\gamma_{\scriptscriptstyle T}\ket\,d\ln r
\end{equation}
(Kaiser \etal 1995; Fahlman \etal 1994)
measures $\overline \kappa(r)$,
the mean surface mass density interior to $r$, minus the mean surface
density in the annulus from $r$ to $r_{\rm max}$, and therefore
provides a lower bound on $\overline \kappa$ and hence on
the mass within an aperture of radius $r$.
Here, the tangential shear is
{$\bra \gamma_{\scriptscriptstyle T}\ket$}
\equals ${1\over 2\pi}\int{\gamma_{\scriptscriptstyle T}\,d\varphi}$, where
$\gamma_{\scriptscriptstyle T}$
\equals$\gamma_1\cos 2\varphi + \gamma_2\sin 2\varphi$, and
$\varphi$ is the azimuthal angle with respect to some chosen center (which
we have taken to be the peak of the smoothed light image in figure 2).

The tangential shear and $\zeta(r)$ are shown for the
various subsamples in figure 7.
A coherent tangential shear pattern is clearly seen in the $I > 21.5$ sample
over a range
of radii from $\sim 50''$ to $\sim 300''$ (though we do not have full
azimuthal coverage for $r > 220''$), and the $\zeta$-statistic shows that the
mean dimensionless surface density rises to $\overline \kappa \simeq 0.25$
at $r \simeq 60''$ with a fractional statistical error of about 20\%.
We calculate the variance in 
$\gamma_\times \equiv - \gamma_1\sin 2\varphi + \gamma_2\cos 2\varphi$.  
If the shear pattern
is circularly symmetric then this should give a fair estimate of
the statistical uncertainty in the shear estimates, and the error
bars in figure 7 are based on this estimate.
For the $I>21.5$ sample for instance, we obtain $\langle \gamma_\times^2
\rangle^{1/2} \simeq 0.6$ which is the value used in the `noise
reconstructions' of figure 4.
The $\gamma$ estimates have uncorrelated statistical uncertainty,
whereas the $\zeta$ estimates are somewhat correlated (as we
have used logarithmically spaced bins in $r$, each $\zeta$-estimate
is just a sum of the $\gamma$ estimates which lie at larger
radii, thus $\zeta$ estimates at small $r$ tend to have errors
which are quite strongly correlated).  We should emphasize that
because we have taken the spatial origin to be the brightest cluster
galaxy, the errors in both $\gamma$ and $\zeta$ are unbiased, and it is
equally likely that we have over- or under-estimated the mass.

The lower panels in figure 7 show graphically
how the distortion strength varies with color and magnitude
of the background objects.  The tangential shear is barely seen
in the bright and red subsamples, while for the faint and blue
samples, $\gamma_T$ lie roughly 30\% higher than the full $I > 21.5$
sample and gives $\overline \kappa (<0.25) \simeq 0.35 \pm 0.07$ and
$\overline \kappa (<0.5) \simeq 0.20 \pm 0.06$.
For the bright and red subsamples the values are $0.13 \pm 0.07$ and
$0.07 \pm 0.05$, and this {\it difference} 
(in shear values between red and blue or bright and faint subsamples)
is significant at the $\simeq 2.2$-sigma
level.  These values are unlikely to have been significantly affected
by cluster contamination, since they only make use of data outside the
aperture.

The average physical surface mass density is obtained by 
multiplying $\overline{\kappa}$ (or $\zeta$) by the critical density,
$\Sigma_{crit}$, and a
lower limit to the total projected mass 
within $r$ is then $M(<r) > \pi\,r^2\,\zeta(r)\,\Sigma_{crit} =
c^2 r^2 \zeta / (4 G D_l \beta)$.
The big uncertainty here is the value for 
$\beta$, 
which varies by a
factor of \approx 5  from $\beta$\approx 0.1 if all the FBGs are at $z_s$\approx
1 to $\beta$\approx 0.5 if the FBGs are at the maximum plausible redshift of
$z_s$\approx 3 (Guhathakurta \etal 1990). 
The critical surface density is $\Sigma_{\rm crit} =
1.95 \times 10^{15} \beta^{-1} M_\odot h {\rm Mpc}^{-2}$ and ranges 
from 
1.7$\times$10$^{16}\,h\,$\msun Mpc$^{-2}$ to
3.9$\times$10$^{15}\,h\,$\msun Mpc$^{-2}$ over this range 
of source redshifts.  If the FBG $N(z)$ shape follows the no evolution
model (as used in Glazebroook et al., 1995) then $\beta \simeq 0.22$
and $\Sigma_{\rm crit} = 8.8 \times 10^{14} M_\odot h {\rm Mpc}^{-2}$.

\begin{figure*}[t]
\vspace{5.0truein}
\includegraphics{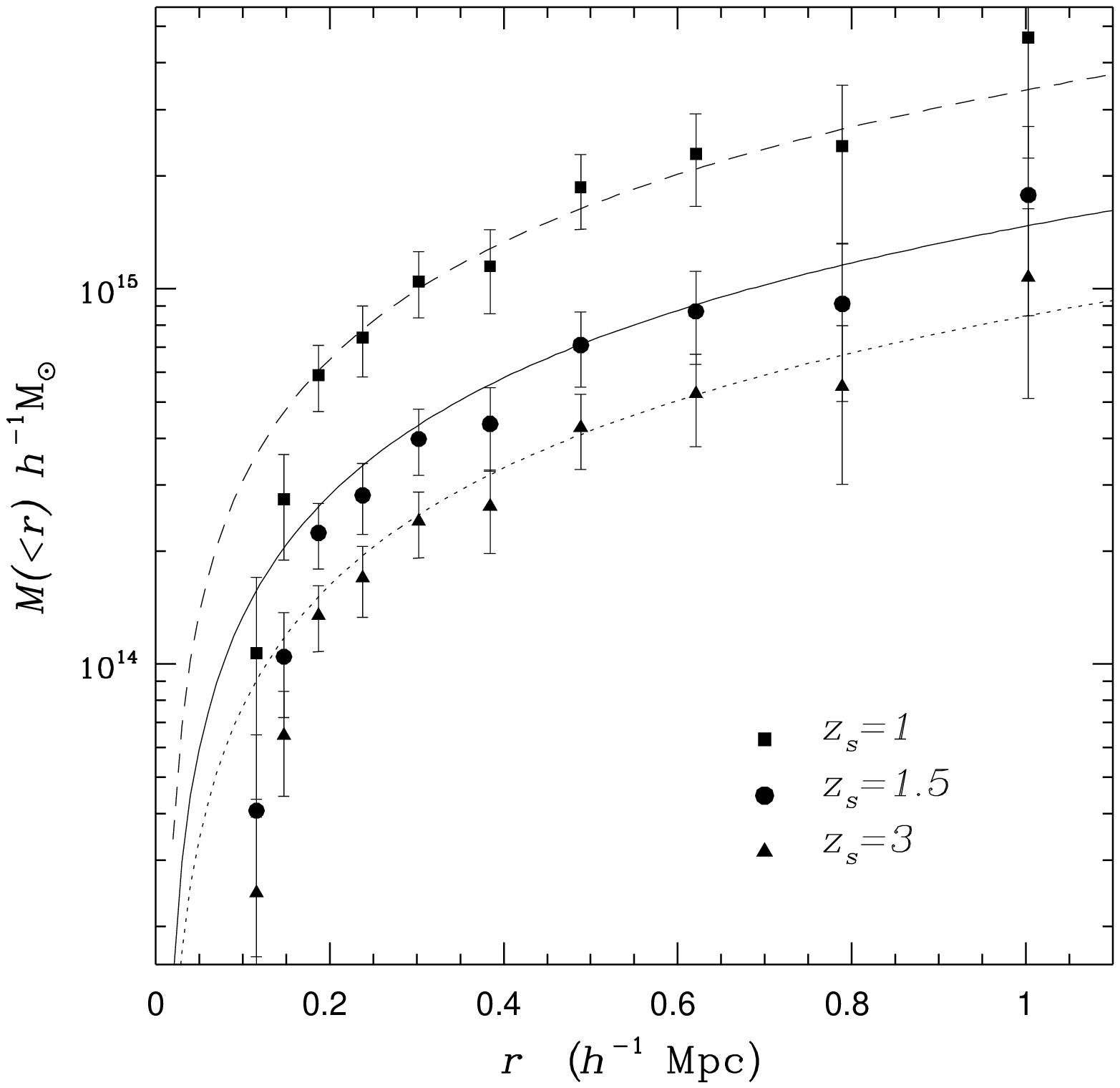}
\begin{center}
\begin{minipage}{6.0in}
\par\noindent
{\small{\sc Figure 8.}
Plot of the radial mass profile
[$M(<r) > \pi\,r^2\,\kappa(r)\,\Sigma_{crit} =
c^2 r^2 {\overline{\kappa}} / (4 G D_l \beta)$]
of \ms1054 using the $\kappa$ (or $\zeta$) values from the $I$\gthan
21.5 sample for three different values
of $\beta$ assuming the faint lensed galaxies lie on sheets at
$z_s$\equals 1,
$z_s$\equals 1.5, and $z_s$\equals 3. The errorbars only reflect the
errors
in $\overline{\kappa}$, and not the uncertainty in $\Sigma_{crit}$.
The dashed, solid and dotted lines are mass profiles for isothermal
spheres with $\sigma$\equals 2200, 1450, and 1100 km/s respectively.
\label{fig8}
}
\end{minipage}
\end{center}
\end{figure*}

In figure 8 we plot the cluster radial 
mass profile for three different
values of $\beta$ corresponding to the faint lensed galaxies lying on
sheets at $z_s$\equals 1, 1.5, and 3. 
Also shown for comparison are isothermal sphere mass
profiles with velocity dispersions 2200, 1450, and 1100 km/s.
A conservative lower bound on the cluster mass is
obtained if we assume that the faint/blue galaxies lie at $z_s = 3$, and
we then find $M(<0.25) = (2.7 \pm 0.6) \times 10^{14} h^{-1} M_\odot$
and $M(<0.5) = (5.9 \pm 1.3) \times 10^{14} h^{-1} M_\odot$.
For the no-evolution $N(z)$, $M(<0.5) = (1.39 \pm 0.29) \times 10^{15} h^{-1} M_\odot$.

We can combine these projected mass estimates with the projected
light estimates of \S3 to obtain the cluster mass-to-light ratio.
Since the mass estimates really measure the mean surface density in the
aperture relative to that in the surrounding annulus we reduce the
luminosity estimates by the expected mean surface brightness (this is
a small correction; roughly 5\% and 15\% for the smaller and larger apertures
respectively).  If we place the faint/blue galaxies at $z_s = 3$ then we
obtain $M/L_V \simeq 250h$ for the small aperture and $M/L_V \simeq 350h$
for the larger (with $\simeq 21$\% statistical uncertainty).
If instead they lie at $z_s = 1.5$, then the mass increases by roughly
70\% and the mass-to-light ratio (for the 0.5$h^{-1}$Mpc aperture)
rises to $M/L_V \simeq 580$. For the no evolution $N(z)$ we find
$M/L_V = (790 \pm 170)h$ and for $z_s < 1$ we would require
$M/L_V > 1600 h$.

Finally, the net shear (which is sensitive to structures outside the
beam) is $\gamma = \{0.019, -0.016\} \pm 0.012$, which is essentially
a null detection, but at a precision level which is already at about
the level of the expected signal from large-scale structure, so the
prospects for constraining the large-scale mass power spectrum
$P(k)$ with large angle surveys is excellent.  

\par\noindent
{\large\bf 5. Discussion}

These results have implications for both the properties of high-$z$
clusters (and therefore for cosmogonical theory), and for the
$N(z)$ of the FBG's.

Regarding the cluster properties, we have found that the mass-to-light
ratio is $> 350 h$, with the lower limit corresponding to having all
the faint lensed galaxies at $z = 3$.  This must be an underestimate
as some of the galaxies surely lie at lower redshifts. For
a more plausible mean redshift of, say $z_s$\equals 1.5, 
we obtain $M/L \simeq 580h$
(though a somewhat lower value for the central mass-to-light ratio), and
for the no-evolution model $M/L \simeq 800h$.
This is quite large compared to values normally obtained from
the X-ray or virial analysis, but is quite consistent with values measured by
weak lensing for other lower-redshift clusters (Fahlman \etal 1994; Smail
\etal 1995; Tyson \& Fischer 1995; Squires \etal 1995).

The high $M/L$ coupled with the high luminosity of the cluster
makes it very massive indeed --- it has the same projected
surface mass density as a Navarro model (Navarro, Frenk \& White 1995) with rotation
velocity $v_{200}$ in the range 2400--2800 km/s, or as an
isothermal sphere with line of sight velocity dispersion of 1100--2200 km/s (see figure 8).
The existence of large clusters like this at high redshift is problematic for
hierarchical cosmological models like CDM with $\Omega$\equals 1.
While this problem has been recognized for some time
(Evrard 1989; Peebles \etal 1989; Gunn 1990), it has not been
taken too seriously because of the lack of conclusive evidence that
any of the few known high-$z$ clusters were truly massive.  We now have
firm evidence for at least one such system.
Using the Press-Schechter approximation, 
the predicted comoving number density of
10$^{15}$\h1\msun clusters at $z$\approx 0.8 in a standard CDM model
($\sigma_8$\equals 1.1) is at least
an order of magnitude lower than the number density at $z$\equals 0
(Vianna \& Liddle 1995). 
But the existence of only one 
10$^{15}$\h1\msun cluster at $z$\approx 0.8 in the
EMSS survey volume corresponds to a comoving number density of order
$n$\approx 5$\times$10$^{-8}\,h^3$ Mpc$^{-3}$ (Luppino \& Gioia 1995), 
comparable to the ``local'' density 
$n\,$({\footnotesize $M$\gthan 10$^{15}$\h1$\!$\msun}) $\sim 10^{-7}\,h^3$
Mpc$^{-3}$ (White \etal 1993).
In mixed dark matter models, the predicted abundance of massive clusters
drops  even more rapidly with redshift than in standard CDM.

The question of the $N(z)$ for the FBG population has been a matter of
debate for some time.
While some of the faint field galaxy population consists of low-redshift
($z$\lthan 0.5) dwarfs, there remains the possibility that large,
star forming galaxies at $z$\gthan 1 make up a significant fraction of
the FBG excess counts, especially at faint magnitudes (Cowie \etal
1995).  There have been hints of this high redshift
component to the FBGs from lensing observations 
of lower redshift ($z$\lthan 0.5) clusters (Fort \etal 1992;  
Kneib \etal 1994), and 
Smail \& Dickinson (1995) have reported the
detection of weak shear by a putative cluster
surrounding the radio galaxy 3C324 at $z$\equals 1.2. Furthermore, 
there is some weak lensing evidence for 
a $z$\approx 1.5 mass concentration coincident with a group of very
faint galaxies that may be  partly responsible for
the lensing of Q2345$+$007 (Mellier \etal 1994; Fischer \etal 1994). 
On the other hand, as mentioned earlier, the failure of
Smail \etal to detect lensing in Cl\thinspace 1603+43 might lead one to the
opposite conclusion.
Our observation shows unequivocally that the 
lensed, faint background galaxies are predominantly blue, and that the
majority of these in the range $23.5 < I < 25.5$ lie at redshifts of
order unity or greater.
Unfortunately we cannot be more precise without some independent
estimate of the mass of the cluster.  What we {\it can} say, however,
is that either extreme case is very interesting.  On one hand, if the
cluster has a mass-to-light ratio at the lower limit of \approx 350$h$,
then nearly all of the FBG's must lie at very high redshift. On the other hand,
to accomodate a more reasonable $N(z)$, such as a `no-evolution' 
model, requires a mass-to-light ratio of \approx 800$h$ and the cluster would
then be exceptionally massive and should have an enormous velocity dispersion 
and X-ray temperature 
(at least in so far as the cluster is approximately spherical and relaxed). 

It is clear, however, that detailed information on the FBG
$N(z)$ is quite within reach.  What is needed is a sample of five or ten
massive clusters at similar redshift to \ms1054, along with a reasonably
complete spectroscopic sample to say $I = 23$.  Although, as we have seen,
it is difficult to detect the lensing in the brighter galaxies, with
a number of lenses the statistics will improve and we should be
able to determine the relative distances for the faint galaxies
relative to the brighter ones, and then use the spectroscopic redshifts to
tie down the overall scale.
Ongoing spectroscopic surveys with the largest telescopes are now beginning
to obtain spectra at the magnitude limits required here. 
Using the Keck Telescope, Cowie \etal (1996) have taken spectra of a sample
of several hundred galaxies nearly complete to $I$\equals 23 ($K$\equals 20,
$B$\equals 24.5). Interestingly, when they split their sample by color 
(at $B-I$\equals 1.6), they find that the blue galaxies divide into distinctly
separate low redshift ($z$\approx 0.25) and high redshift ($z$\gthan 0.8) 
populations, with the bulk of the faintest blue galaxies located at high redshift
(see figs. 18 and 20 in Cowie \etal 1996). 
Combining these observations with weak lensing, it should
be possible to constrain the redshifts
of galaxies that are several magnitudes fainter than 
will be accessible to spectroscopy even with 8--10\thinspace m
telescopes in the forseeable future.

It is a pleasure to thank Lev Kofman, Isabella Gioia, Ken Chambers, Doug Clowe, Megan Donahue,
Mark Metzger, Karl Glazebrook, Neal Trentham and Len Cowie for stimulation, help and advice.

\clearpage

\begin{figure*}[p]
\vspace{9.0truein}
\includegraphics{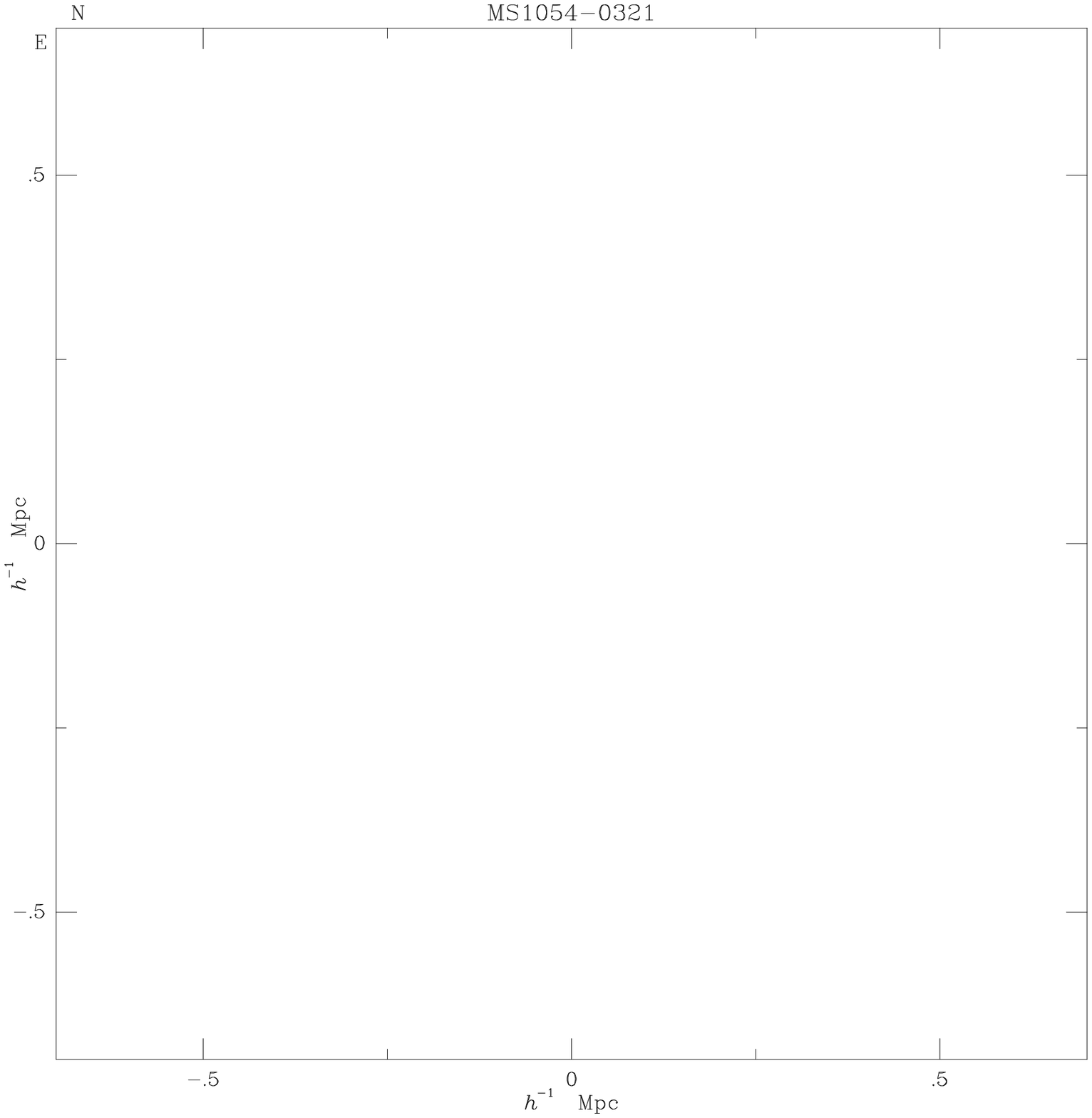}
\vspace{-0.8in}
\begin{center}
\begin{minipage}{7.0in}
{\small{\sc Figure 1 [Plate 1].}
True color image of MS\thinspace 1054$-$0321 formed from the $B$, $R$,
and $I$ CCD frames. This image measures 1536 $\times$ 1536  pixels and
covers a field of $5'.6\times5'.6$ (1.4$h^{-1}\times$1.4$h^{-1}$ Mpc
at $z$\equals 0.83).
}
\end{minipage}
\end{center}
\end{figure*}

\begin{figure*}[p]
\vspace{9.0truein}
\includegraphics{fig3.ps}
\vspace{-1.0truein}
\par\noindent
\begin{center}
\begin{minipage}{7.0in}
{\small{\sc Figure 3 [Plate 2].}
Full 2048$\times$2048 pixel $I$-band CCD image of \ms1054 with the
ellipses drawn around all the 2395 objects in the $I$\gthan 21.5 catalog.
}
\end{minipage}
\end{center}
\end{figure*}

\begin{figure*}[p]
\vspace{8.0truein}
\includegraphics{fig5.ps}
\begin{center}
\begin{minipage}{6.5in}
\par\noindent
{\small{\sc Figure 5 [Plate 3].}
Contour plot of the surface mass density (black contour lines) and
cluster light distribution (white contour lines) overlaid on the
2048$^2$ pixel optical image of the cluster.
Both the mass contours and the light contours
have been smoothed with a gaussian of scale length 0$''$.35.
The image measures 1.86\h1$\times$1.86\h1{Mpc}, and
an $r$\equals0.5\h1{Mpc} circle centered on the BCG  is shown for reference.
}
\end{minipage}
\end{center}
\end{figure*}

\end{document}